\documentclass[twovoluments,showpacs,preprintnumbers]{revtex4}
\usepackage{amssymb}
\usepackage{amsmath}
\usepackage{txfonts}
\usepackage[mathscr]{eucal}

\input{tcilatex}
\begin{document}

\title{Hawking radiation via Landauer transport model}
\author{Xiao-Xiong Zeng}
\email{xxzengphysics@163.com}
\affiliation{Department of Physics, Institute of Theoretical Physics, Beijing Normal
University, Beijing, 100875, China}
\author{Wen-Biao Liu (corresponding author)}
\email{wbliu@bnu.edu.cn}
\affiliation{Department of Physics, Institute of Theoretical Physics, Beijing Normal
University, Beijing, 100875, China}

\begin{abstract}
Recently, Nation et al confirmed that fluxes of Hawking radiation energy and
entropy from a black hole can be regarded as a one-dimensional (1D) Landauer
transport process. Their work can be extended to background spacetimes with
gauge potential. The result shows that the energy flux, which is indicated
to be equal to the energy-momentum tensor flux, contains not only the
contribution of thermal flux but also that of particle flux. We find that
the charge can also be regarded as transporting via a 1D quantum channel and
the charge flux is equal to the gauge flux.

\end{abstract}

\maketitle

\section{Introduction}

In 1974, Hawking \cite{1} published his striking argument in which black
holes are not black but can radiate in the curved spacetimes background by
the calculation of Bogoliubov transformation coefficients connecting an
initial vacuum far outside the collapsing matter to a final vacuum after the
black hole formed. It provides not only a firm foundation to black hole
thermodynamics but also a platform to research the quantum gravity. Until
now, there are many methods to derive Hawking radiation except that of
Hawking, such as path integral \cite{2,2-1}, trace anomaly \cite{3} or
gravitational anomaly \cite{4,4-1,4-2,4-3,4-4,4-5,4-6,4-7,4-8,4-9}, periodic
Green function \cite{5,5-1}, quantum tunneling \cite%
{6,6-1,6-2,6-3,6-4,6-5,6-6,6-7,6-8,6-9,6-10}, holographic principle \cite%
{7,7-1}, and so on. All of them confirmed a fact that Hawking radiation is
the intrinsic property of the event horizon.

According to the initial description of Hawking, the radiation arises from
the production of virtual particle pairs spontaneously near the horizon due
to the vacuum fluctuation. When the negative energy virtual particle tunnels
inwards, the positive energy virtual particle materializes as a real
particle and escapes to infinity. However, how does the positive energy
particle run? Recently, Nation et al \cite{8} proposed that the positive
energy particle escaped to infinity via a 1D quantum channel by making use
of the Landauer transport model. They found that the flux of energy was
equal to Hawking radiation flux perfectly for both fermions and bosons,
indicating that the energy flow is particle statistics independent.

The Landauer transport model was first proposed to study electrical
transport in mesoscopic circuits \cite{9,9-1}, and subsequently used to
describe thermal transport \cite{10,10-1,10-2,10-3}. In 2000, the phononic
quantized thermal conductance counterpart was measured for the first time 
\cite{11}. For the 1D quantum channel of thermal conductance, it is supposed
that there are two thermal reservoirs, which are coupled adiabatically
through an effective 1D connection, characterized by temperatures $T_{L}$, $%
T_{R}$ and chemical potentials $u_{L}$, $u_{R}$, where subscripts $L$ and $R$
denotes the left and right thermal reservoirs respectively. Because of the
temperature difference, the thermal transportation will happen.

The key idea in the work of Nation et al is that the black hole and its
environment were viewed as two thermal reservoirs. As temperature in the
right thermal reservoir, namely the environment of a black hole, approaches
to zero and the scattering effect can be ignored, the energy flow can be
obtained using the Landauer transport model, which was shown to be equal to
the energy-momentum tensor expectation value of an infinite observer. They
also found the energy current and the upper limit of entropy current were
same for both bosons and fermions.

In this paper, we intend to extend their work to background spacetimes with
chemical potential. We will take the higher dimensional Reissner-Nordstr$%
\ddot{o}$m-de Sitter black hole, Kerr-Newman black hole, and 5D black ring
as examples to check whether this model is still valid for different
geometry. For spacetimes with chemical potential, when the near horizon
conformal symmetry is considered, the expectation value of charge flow
should be investigated except the flow of energy-momentum tensor. Thus, when
the Landauer transport model is researched, one should also explore whether
the energy flow contains the contribution of particle flow besides the flux
of thermal radiation. It is valuable to explore the relation between the
charge flow and gauge flux too.

The remainder of this paper is arranged as follows. In Sec.II, we review
Hawking radiation flux and gauge flux from a 2D conformal metric by
calculating vacuum expectation values of energy-momentum tensor. In Sec.III,
flows of energy and charge of both fermions and bosons are studied from the
viewpoint of 1D quantum transport. In Sec.IV, Hawking radiation flux and
charge flux from the higher dimensional Reissner-Nordstr$\ddot{o}$m-de
Sitter black hole, Kerr-Newman black hole, and 5D black ring are
investigated. Finally, our discussion and conclusion are given in Sec.V.
Throughout this paper, we use the units $G=\hbar =c=k_{B}=1$.

\section{Hawking radiation flux}

For Hawking radiation flux in the spacetimes with horizon, there are several
methods to derive it. In the following, we will review the idea of Unruh 
\cite{12} to calculate the vacuum expectation values of energy-momentum
tensor. For any background spacetimes, with nonvanishing chemical potential $%
A_{t}$ in the gauge field, one may reduce it to the 2D form as 
\begin{equation}
ds^{2}=-f(r)dt^{2}+f(r)^{-1}dr^{2},  \label{1}
\end{equation}%
where the event horizon satisfies $f(r_{h})=0$. For this metric, one can
define the null coordinates $u=t+r^{\ast }$, $v=t-r^{\ast }$, by introducing
the tortoise coordinate transformation defined as $dr^{\ast }=\frac{1}{f(r)}%
dr$, and Kruskal coordinates $U=-\frac{1}{\kappa }e^{-{\kappa u}}$, $V=\frac{%
1}{\kappa }e^{\kappa v}$, which can induce the corresponding conformal form,
respectively%
\begin{equation}
ds^{2}=-f(r)dudv,  \label{2}
\end{equation}%
\begin{equation}
ds^{2}=-f(r)e^{-2{\kappa r^{\ast }}}dUdV,  \label{3}
\end{equation}%
where $\kappa =\frac{1}{2}\frac{\partial {f(r)}}{\partial r}\mid _{r_{h}}$
is the surface gravity.

It is well-known that the classical Einstein field equation can be derived
from the classical action by the minimal variational principle. In the
semiclassical theory, this principle is still valid, namely%
\begin{equation}
\frac{2}{\sqrt{-g}}\frac{\delta \Gamma }{\delta g^{\mu \nu }}=\langle T_{\mu
\nu }\rangle ,  \label{4}
\end{equation}%
in which $\Gamma $ is the effective action with central charge $c=1$.
Obviously, one should first give the action in order to get the expectation
values of energy-momentum tensor. In the gravitational field with $U(1)$
gauge field background, the energy-momentum tensor are solved as \cite%
{13,14,14-1,14-2,14-3,14-4}%
\begin{eqnarray}
\langle T_{\mu \nu }\rangle &=&\frac{1}{48\pi }[{g_{\mu \nu }(2R-\frac{1}{2}%
\nabla ^{\rho }S\nabla _{\rho }S})-2\nabla _{\mu }\nabla _{\nu }S  \notag \\
&&+\nabla _{\mu }S\nabla _{\nu }S]+\frac{e^{2}}{\pi }({\nabla _{\mu }B\nabla
_{\nu }B-\frac{1}{2}g_{\mu \nu }\nabla ^{\rho }B\nabla _{\rho }B}),
\label{5}
\end{eqnarray}%
\begin{equation}
\langle J^{\mu }\rangle =\frac{e^{2}}{\pi }\frac{1}{\sqrt{-g}}\epsilon ^{\mu
\nu }\partial _{\nu }B,  \label{6}
\end{equation}%
in which $R$ is the 2D scalar curvature, $\Delta _{g}$ is the Laplacian, $%
\epsilon ^{\mu \nu }$ represents the 2D Levi-Civita tensor, and%
\begin{equation}
S(x)=\int {d^{2}y\frac{1}{\Delta _{g}}(x,y)\sqrt{-g}R(y)},  \label{7}
\end{equation}%
\begin{equation}
B(x)=\int {d^{2}y\frac{1}{\Delta _{g}}(x,y)\epsilon ^{\mu \nu }\partial
_{\mu }A_{\nu }(y)}.  \label{8}
\end{equation}

The above equations are introduced by Iso et al \cite{13} recently to get
the Hawking radiation flux from the viewpoint of anomaly. In order to get
the last values of Eqs.(\ref{5}) and (\ref{6}), the vacuum states should be
considered. Usually, there are three kinds of vacuum states. Boulware vacuum
corresponds to our familar empty state for large radius outside of the black
hole, the renormalized physical quantities are divergent. Hartle-Hawking
vacuum corresponds to a black hole in unstable equilibrium with infinite sea
of black body radiation. Therefore both of Boulware vacuum and
Hartle-Hawking vacuum can't describe the ingoing and outgoing modes
felicitously. However this problem can be overcome by the Unruh vacuum, in
which the ingoing vacuum is defined with respect to the advanced null
coordinate $\frac{\partial }{\partial v}$ and the outgoing vacuum is defined
with respect to the Kruskal coordinate $\frac{\partial }{\partial U}$. That
is to say, in the advanced null coordinate, there is a finite mount of flux
at the horizon while no flux at infinity. Meanwhile, there is no flux at the
horizon while a finite mount of flux at infinity in the Kruskal coordinate.
As the relations $J_{U}=-\frac{J_{u}}{\kappa U}$ and $T_{UU}=(\frac{1}{%
\kappa U})^{2}T_{uu}$ between the retarded null coordinate and Kruskal
coordinate are considered, the gauge and Hawking radiation fluxes can be,
respectively, written as [39]%
\begin{equation}
\langle J_{u}\rangle =\frac{e^{2}}{2\pi }[A_{t}(r)-A_{t}(r_{h})],  \label{13}
\end{equation}%
\begin{equation}
\langle J_{v}\rangle =\frac{e^{2}}{2\pi }A_{t}(r),  \label{14}
\end{equation}%
\begin{equation}
\langle T_{uu}\rangle =\frac{1}{192\pi }(-f^{\prime 2}+2ff^{\prime \prime })+%
\frac{e^{2}}{4\pi }[A_{t}(r)-A_{t}(r_{h})]^{2}+\frac{f^{2}(r_{h})}{192\pi },
\label{15}
\end{equation}%
\begin{equation}
\langle T_{vv}\rangle =\frac{1}{192\pi }(-f^{\prime 2}+2ff^{\prime \prime })+%
\frac{e^{2}}{4\pi }A_{t}^{2}(r).  \label{16}
\end{equation}%
From Eq.(\ref{13}) and Eq.(\ref{14}), we find the infinite observers will
observe a mount of gauge flux $-\frac{e^{2}}{2\pi }A_{t}(r_{h})$ while the
free falling observers will see a negative flow $\frac{e^{2}}{2\pi }%
A_{t}(r_{h})$ at the event horizon. Similarly, Eq.(\ref{15}) indicates that
the infinite observers will see a bunch of Hawking radiation flow $\frac{\pi 
{T_{h}^{2}}}{12}+\frac{e^{2}}{4\pi }A_{t}^{2}(r_{h})$, in which $T_{h}=\frac{%
\kappa }{2\pi }$ is Hawking temperature, and Eq.(\ref{16}) indicates the
free falling observers will see that a bunch of Hawking temperature flow $-%
\frac{\pi {T_{h}^{2}}}{12}+\frac{e^{2}}{4\pi }A_{t}^{2}(r_{h})$ drops into
the event horizon.

\section{Landauer transport model}

In 1957, Rolf Landauer presented a very intuitive interpretation of electron
conduction from the viewpoint of 1D quantum transport \cite{15,15-1}. This
theory was then extended to the case of quantum thermal transport, where two
infinite reservoirs with temperature $T_{L}$, $T_{R}$ and chemical potential 
$u_{L}$, $u_{R}$ respectively are adiabatically connected to each other
through a 1D channel. Universally, there are several different distribution
functions to describe the thermal equilibrium distribution of particles in
reservoir, here we adopt the Haldane's statistics \cite{16}%
\begin{equation}
f_{g}(E)=\{ \omega \lbrack \frac{(E-u)}{T}]+g\}^{-1},  \label{17}
\end{equation}%
where the function $\omega (x)$ satisfies the relation%
\begin{equation}
\omega (x)^{g}[1+\omega (x)]^{1-g}=e^{x},  \label{18}
\end{equation}%
in which $g=0$, $g=1$ correspond bosons and fermions respectively. In fact,
fermions and bosons are the special cases of this statistics theory, it has
shown the value of $g$ also can be taken as $1/4$, $1/3$, $1/2$, $2$, $3$, $%
4 $ \cite{17}.

For the sake of simplicity in the present investigation, we restrict
ourselves to ballistic transport, which means that the channel currents do
not interfere with each other \cite{8}. The left (right) components of the
single channel energy current hence can be written as%
\begin{equation}
\dot{E}_{L(R)}=\frac{T_{L(R)}^{2}}{2\pi }\int_{x_{L(R)}^{0}}^{\infty }dx(x+%
\frac{u_{L(R)}}{T_{L(R)}})f_{g}(x),  \label{19}
\end{equation}%
in which $x_{L(R)}^{0}=-\frac{u_{L(R)}}{T_{L(R)}}$. As done in Ref.\cite{8},
we define the zero of energy with respect to the longitudinal component of
the kinetic energy. The total energy current is then just $\dot{E}=\dot{E}%
_{L}-\dot{E}_{R}$.

We first consider the case of fermions, where the contributions of
antiparticles should be considered. According to the viewpoint of Landauer,
the maximum energy flow of fermions can be treated as the combination of
fermionic particle and antiparticle single channel currents. In this case,
as $x=\frac{(E-u)}{T}$, $y=\frac{(E+u)}{T}$ are defined, the flow of energy
can be expressed as%
\begin{equation}
\dot{E}_{L(R)}=\frac{T_{L(R)}^{2}}{2\pi }\bigg[\int_{\frac{-u_{L(R)}}{%
T_{L(R)}}}^{\infty }dx(x+\frac{u_{L(R)}}{T_{L(R)}})\frac{1}{e^{x}+1}+\int_{%
\frac{u_{L(R)}}{T_{L(R)}}}^{\infty }dy(y+\frac{u_{L(R)}}{T_{L(R)}})\frac{1}{%
e^{y}+1}\bigg].  \label{20}
\end{equation}%
Due to the existence of chemical potential, the lower limit of the integral
isn't zero, so one can not finish the integral directly. To get the final
value, we first vary the lower limit of integral, that is%
\begin{eqnarray}
\dot{E}_{L(R)} &=&\frac{T_{L(R)}^{2}}{2\pi }\bigg[\int_{0}^{\infty }dx(x+%
\frac{u_{L(R)}}{T_{L(R)}})\frac{1}{e^{x}+1}+\int_{0}^{\frac{u_{L(R)}}{%
T_{L(R)}}}dx(-x+\frac{u_{L(R)}}{T_{L(R)}})\frac{1}{e^{-x}+1}+  \notag \\
&&\int_{0}^{\infty }dy(y+\frac{u_{L(R)}}{T_{L(R)}})\frac{1}{e^{y}+1}%
-\int_{0}^{\frac{u_{L(R)}}{T_{L(R)}}}dy(y+\frac{u_{L(R)}}{T_{L(R)}})\frac{1}{%
e^{y}+1}\bigg].  \label{21}
\end{eqnarray}%
As $\frac{1}{e^{-x}+1}=1-\frac{1}{e^{x}+1}$ is replaced, Eq.(\ref{21}) takes
the form as%
\begin{eqnarray}
\dot{E}_{L(R)} &=&\frac{T_{L(R)}^{2}}{2\pi }\bigg[\int_{0}^{\infty }\frac{x}{%
e^{x}+1}dx+\int_{0}^{\infty }\frac{y}{e^{y}+1}dy+2\frac{u_{L(R)}}{T_{L(R)}}%
(\int_{0}^{\infty }\frac{1}{e^{x}+1}dx  \notag \\
&&-\int_{0}^{\frac{u_{L(R)}}{T_{L(R)}}}\frac{1}{e^{x}+1}dx)+\frac{%
u_{L(R)}^{2}}{2T_{L(R)}^{2}}\bigg].  \label{22}
\end{eqnarray}

According to the technique of Landau \cite{18}, the convergence rate of $%
\frac{1}{e^{x}+1}$ is very fast, the upper limit of integral $\frac{u_{L(R)}%
}{T_{L(R)}}$ therefore can be changed into infinity because $\frac{u_{L(R)}}{%
T_{L(R)}}>1$. The energy current now can be expressed as%
\begin{eqnarray}
\dot{E}_{L(R)} &=&\frac{T_{L(R)}^{2}}{2\pi }\bigg[\int_{0}^{\infty }\frac{x}{%
e^{x}+1}dx+\int_{0}^{\infty }\frac{y}{e^{y}+1}dy+\frac{u_{L(R)}^{2}}{%
2T_{L(R)}^{2}}\bigg].  \label{22}
\end{eqnarray}
Finishing the integration, we have 
\begin{equation}
\dot{E}=\dot{E}_{L}-\dot{E}_{R}=\frac{\pi }{12}({T_{L}^{2}-T_{R}^{2}})+\frac{%
1}{4\pi }({u_{L}^{2}-u_{R}^{2}}).  \label{23}
\end{equation}%
Eq.(\ref{23}) tells us that the energy current flowing through the 1D system
can be divided into two independent components: the flux of particles, which
is only related to the particles, and the flux of thermal radiation, which
is determined by the temperature of the reservoirs entirely. Our result,
obviously, is different from that of Nation et al because of the emergence
of chemical potential.

According to the Landauer transport theory, the charge can also be
transported by the 1D quantum channel. As the scattering effect is ignored,
the charge flux can be expressed as%
\begin{equation}
\dot{I}=\frac{T_{L(R)}e}{2\pi }\int_{-\frac{u_{L(R)}}{T_{L(R)}}}^{\infty }%
\frac{1}{{e^{x}}+1}dx.  \label{24}
\end{equation}%
For the case of fermions, the contribution of antiparticle should also be
considered as%
\begin{equation}
\dot{I}=\frac{T_{L(R)}e}{2\pi }\int_{-\frac{u_{h}}{T_{L(R)}}}^{\infty }\frac{%
1}{{e^{x}}+1}dx+\frac{T_{L(R)}e}{2\pi }\int_{\frac{u_{L(R)}}{T_{L(R)}}%
}^{\infty }\frac{1}{{e^{y}}+1}dy.  \label{25}
\end{equation}%
As the same case in Eq.(\ref{22}), when the lower limit of integral $\frac{%
u_{L(R)}}{T_{L(R)}}$ is changed as infinity, the second term in the right of
Eq.(\ref{25}) will vanish and the first term will lead to%
\begin{equation}
\dot{I}=\frac{e}{2\pi }(u_{L}-u_{R}),  \label{26}
\end{equation}%
which shows that the charge current doesn't depend on the temperature and it
only relates to the charge and chemical potential.

Now, we check whether the Landauer transport model is valid for the bosons.
Putting $g=0$ into Eq.(\ref{19}), the energy current can be expressed as%
\begin{equation}
\dot{E}_{L(R)}=\frac{T_{L(R)}^{2}}{2\pi }\int_{\frac{-u_{L(R)}}{T_{L(R)}}%
}^{\infty }dx(x+\frac{u_{L(R)}}{T_{L(R)}})\frac{1}{e^{x}-1}.  \label{27}
\end{equation}

After varying the lower limit of integral and adopting $\frac{1}{e^{-x}-1}%
=-1-\frac{1}{e^{x}-1}$, the above equation can be rewritten as 
\begin{equation}
\dot{E}_{L(R)}=\frac{T_{L(R)}^{2}}{2\pi }\bigg[\int_{0}^{\infty }(x+\frac{%
u_{L(R)}}{T_{L(R)}})\frac{1}{e^{x}-1}dx-\int_{0}^{\frac{u_{L(R)}}{T_{L(R)}}}(%
{\frac{u_{L(R)}}{T_{L(R)}}-x})\frac{1}{e^{x}-1}dx-\frac{u_{L(R)}^{2}}{%
2T_{L(R)}^{2}}\bigg].  \label{28}
\end{equation}%
Adopting the similar techniques as the case of fermions, we get 
\begin{equation}
\dot{E}=\dot{E}_{L}-\dot{E}_{R}=\frac{\pi }{12}({T_{L}^{2}-T_{R}^{2}})+\frac{%
1}{4\pi }({u_{L}^{2}-u_{R}^{2}}),  \label{29}
\end{equation}%
which is consistent with that of fermions.

In the case of system with chemical potential, we have got energy fluxes for
both fermions and bosons, which contains the contributions of particles and
thermal radiation. We find that they are more complicated, but are still
independent on the kinds of particles.

\section{Hawking radiation from black holes via Landauer transport model}

In this section, we discuss Hawking radiation of black holes via the
Landauer transport model. For this model, one necessity is a heat source.
Recall that any spacetimes with horizon will emit Hawking radiation with
temperature $\frac{\kappa}{2\pi}$, where $\kappa$ is surface gravity, so the
black hole can be treated as one heat source with Hawking temperature on the
horizon. When the black hole and its surrounding, which have temperature $%
T_{L}=T_{h},\quad T_{R}=0$ and chemical potential $u_{L}=u_{h},\quad u_{R}=0$
respectively, are regarded as thermal reservoirs connected by a 1D quantum
tunnel, we find the energy flux and charge flux can be expressed as%
\begin{equation}
\dot{E}=\dot{E}_{L}-\dot{E}_{R}=\frac{\pi T_{h}^{2}}{12}+\frac{u_{h}^{2}}{%
4\pi },  \label{31}
\end{equation}%
\begin{equation}
\dot{I}=\frac{eu_{h}}{2\pi },  \label{32}
\end{equation}%
where we have defined the chemical potential $u_{h}=-eA_{t}(h)$. Based on
above formulisms, we take different background spacetimes as examples to
check whether the energy flux and charge flux are equal to the Hawking
radiation flux and gauge flux next.

\subsection{Higher dimensional Reissner-Nordstr$\ddot{o}$m-de Sitter black
hole}

The metric of the higher dimensional Reissner-Nordstr$\ddot{o}$m-de Sitter
black hole with a positive cosmological constant $\Lambda =n(n+1)/2l^{2}$
takes the form as \cite{19}%
\begin{equation}
ds^{2}=-f(r)dt^{2}+f(r)^{-1}dr^{2}+r^{2}d\Omega _{n}^{2},  \label{33}
\end{equation}%
in which%
\begin{equation}
f(r)=1-\frac{\omega _{n}M}{r_{n-1}}+\frac{\omega _{n}Q^{2}}{%
2(n-1)V_{n}r^{2n-2}}-\frac{r^{2}}{l^{2}},\omega _{n}=\frac{16\pi }{nV_{n}},
\label{34}
\end{equation}%
where $M$ and $Q$ are the mass and the electric charge of the black hole, $l$
and $V_{n}$ denotes the curvature radius of the de Sitter space and the
volume of $d\Omega _{n}^{2}$, which represents a $n-$dimensional spherical
hyper-surface of unit radius.

For this spacetime, we can obtain the horizons from $f(r)=0$. Usually, there
are three positive roots, which satisfy $r_{-}<r_{h}<r_{c}$, and a negative
root for $n\geq 2$, where $r_{-},$ $r_{h},$ $r_{c}$ stands for the inner
horizon, event horizon, and cosmological horizon respectively. In principle,
Hawking radiation can be emitted from both event horizon and cosmological
horizon. However, the radiation from cosmological horizon is very weak and
the calculation there is similar to the case of event horizon, so we only
consider radiation from the event horizon, namely only the event horizon is
treated as the thermal source here.

The electric-magnetic gauge potential of the black hole is%
\begin{equation}
A_{t}=-\frac{Q}{{(n-1)}V_{n}r^{n-1}}.  \label{35}
\end{equation}%
On the base of the definition of surface gravity, we can also get the
Hawking temperature at the event horizon%
\begin{equation}
T_{h}=\frac{\kappa }{2\pi }=\frac{1}{2\pi }[\frac{(n-1)\omega _{n}M}{%
2r_{h}^{n}}-\frac{\omega _{n}Q^{2}}{2V_{n}r_{h}^{(2n-1)}}-\frac{r}{l^{2}}].
\label{36}
\end{equation}

For the Higher dimensional Reissner-Nordstr$\ddot{o}$m-de Sitter black hole,
we also can reduce it to the 2D metric by dimensional reduction techniques
as done in Ref.[39] effectively. Then by the similar skills as in Sec.II, we
can get the fluxes of Hawking radiation and charge, which agree with the
results in Eqs.(9-12).

From the Landauer transport model, when the event horizon of Higher
dimensional Reissner-Nordstr$\ddot{o}$m-de Sitter black hole is regarded as
the heat source, we can get the energy flux and charge flux 
\begin{equation}
\dot{E}=\frac{\pi T_{h}^{2}}{12}+\frac{e^{2}A_{t}^{2}(r_{h})}{4\pi },
\label{37}
\end{equation}%
\begin{equation}
\dot{I}=\frac{e^{2}Q}{2\pi (n-1)V_{n}r_{h}^{n-1}},  \label{38}
\end{equation}%
in which the chemical potential $u_{h}=-eA_{t}(h)$ is used. Comparing them
with Eq.(\ref{15}) and Eq.(\ref{13}), we find the energy flux and charge
flux are equal to the Hawking radiation flux and gauge flux completely,
namely thermal radiation and charge can be transported by a 1D quantum
channel. As $n=2$ or $n=2,$ $l\rightarrow \infty $, one can also get Hawking
radiation flux and charge flux of 4D Reissner-Nordstr$\ddot{o}$m-de Sitter
black hole or 4D Reissner-Nordstr$\ddot{o}$m black hole from the viewpoint
of 1D quantum transport.

\subsection{Kerr-Newman black hole}

As analyzed in Ref.[39], the rotating and charged black hole owns not only
the $U(1)$ gauge symmetry with electromagnetic field but also the induced $%
U(1)$ gauge symmetry associated with isometry along the $\phi $ direction,
so one should consider the contributions of both electric charge $e$ and
azimuthal quantum number $m$ when we calculate the energy-momentum tensor
flux and charge flux. In Boyer-Lindquist coordinates, the 4D Kerr-Newman
black hole can be expressed as \cite{20,20-1}%
\begin{equation}
ds^{2}=-\frac{\Delta }{\rho ^{2}}\left( {dt-a\sin ^{2}\theta d\phi }\right)
^{2}+\frac{\rho ^{2}}{\Delta }\left( {dr^{2}+\Delta d\theta ^{2}}\right) +%
\frac{\sin ^{2}\theta }{\rho ^{2}}\left[ {adt-\left( {r^{2}+a^{2}}\right)
d\phi }\right] ^{2},  \label{39}
\end{equation}%
with $\rho ^{2}=r^{2}+a^{2}\cos ^{2}\theta $, $\Delta
=r^{2}-2Mr+a^{2}+Q^{2}=(r-r_{h})(r-r_{-})$, where $M$, $Q$, $a$ are the
black hole mass, charge and the angular momentum per unit mass, $r_{h}=r_{+}$
and $r_{-}$ denote the outer and inner horizons, which are%
\begin{equation*}
r_{\pm }=M\pm \sqrt{M_{2}-Q_{2}-a_{2}}.
\end{equation*}%
The background gauge field is%
\begin{equation}
A=\frac{Qr}{r^{2}+a^{2}cos^{2}\theta }{(dt-sin^{2}{\theta }d\phi )}.
\label{40}
\end{equation}%
The Hawking temperature can be calculated as%
\begin{equation}
\quad T_{h}=\frac{r_{h}-r_{-}}{4\pi (r_{h}^{2}+a^{2})}.  \label{43}
\end{equation}%
To derive Hawking radiation from a Kerr-Newman black hole via anomalous
point of view, Ref.\cite{13} have given the effective 2D metric with
nonvanishing metric component%
\begin{equation}
g_{tt}=-f(r)=-\frac{\Delta }{r^{2}+a^{2}},g_{rr}=f(r)^{-1},  \label{41}
\end{equation}%
and the corresponding $U(1)$ gauge background%
\begin{equation}
A_{t}=-\frac{Qr}{r^{2}+a^{2}}-\frac{a}{r^{2}+a^{2}}.  \label{42}
\end{equation}

Based on above equations, we also can get the energy-momentum tensor flux
and gauge flux by the similar skills as in Sec.II. Note that here, the
fluxes of energy-momentum tensor and charge depends not only the electric
charge but also the azimuthal quantum number $m$ because of the difference
of spacetime geometry.

Making use of the Landauer transport model, we can also get the energy flux
and charge flux from the Kerr-Newman black hole%
\begin{equation}
\dot{E}=\frac{\pi T_{h}^{2}}{12}+\frac{1}{4\pi }(\frac{eQr_{h}}{%
r_{h}^{2}+a^{2}}+\frac{ma}{r_{h}^{2}+a^{2}})^{2},  \label{44}
\end{equation}%
\begin{equation}
\dot{I_{m}}=\frac{m}{2\pi }(\frac{eQr_{h}}{r_{h}^{2}+a^{2}}+\frac{ma}{%
r_{h}^{2}+a^{2}}),  \label{45}
\end{equation}%
\begin{equation}
\dot{I_{e}}=\frac{e}{2\pi }(\frac{eQr_{h}}{r_{h}^{2}+a^{2}}+\frac{ma}{%
r_{h}^{2}+a^{2}}),  \label{46}
\end{equation}%
where we have defined the chemical potential as%
\begin{equation}
u_{h}=\frac{eQr_{h}}{r_{h}^{2}+a^{2}}+\frac{ma}{r_{h}^{2}+a^{2}}.  \label{47}
\end{equation}%
Comparing the fluxes with those obtained from conformal symmetry, we find
they are equal too. Apparently, not only thermal radiation and particles,
but also the angular momentum of the Kerr-Newman black hole will be
transported by a 1D quantum channel.

\subsection{5-dimensional black ring}

The 5D neutral black ring, which was first found in \cite{21}, is a vacuum
solution of five-dimensional general relativity. Black ring obeys similar
thermodynamics laws as black holes. However, their horizon topology is not
spherical as black holes but with topology $S1\times S2$. Therefore, it is
also significant to check whether this background spacetime is valid for the
Landauer transport model.

The 5D neutral black ring is \cite{21}%
\begin{eqnarray}
ds^{2} &=&-\frac{F(y)}{F(x)}\left( dt-C(\nu ,\lambda )R\frac{1+y}{F(y)}d\psi
\right) ^{2}  \notag \\
&&+\frac{R^{2}}{(x-y)^{2}}F(x)\left[ -\frac{G(y)}{F(y)}d\psi ^{2}-\frac{%
dy^{2}}{G(y)}+\frac{dx^{2}}{G(x)}+\frac{G(x)}{F(x)}d\varphi ^{2}\right] ,
\label{48}
\end{eqnarray}%
in which%
\begin{eqnarray*}
&&F(\xi )=1+\lambda \xi ,\quad G(\xi )=(1-\xi ^{2})(1+\nu \xi ), \\
&&C(\nu ,\lambda )=\sqrt{\lambda (\lambda -\nu )\frac{1+\lambda }{1-\lambda }%
}.
\end{eqnarray*}%
The parameters $\lambda $ and $\nu $ are dimensionless, which take value in
the range $0<\nu \leq \lambda <1$, and $R$ corresponds roughly to the radius
of the ring circle. The mass and angular momentum of the black ring relates
to these parameters as $M=3\pi R^{2}\lambda /[4(1-\nu )]$, $J=\pi R^{3}\sqrt{%
\lambda (\lambda -\nu )(1+\lambda )}/[2(1-\nu )^{2}]$. The coordinates $\phi 
$ and $\psi $ are two cycles of the black ring, $x$ and $y$ take the range
as $-1\leq x\leq 1$, $-\infty \leq y\leq -1$. The horizon is located at $%
y=y_{h}=-1/\nu $.

After the consideration of dimensional reduction, the 2D metric, with gauge
charge $l$ and $U(1)$ gauge potential%
\begin{equation}
A_{t}(y)=\frac{F(y)}{CR{(1+y)}},  \label{49}
\end{equation}%
can be expressed as%
\begin{equation}
ds^{2}=-f(y)dt^{2}+{f(y)}^{-1}dy^{2},  \label{50}
\end{equation}%
in which%
\begin{equation}
f(y)=\frac{\sqrt{-F(y)}}{CR{(1+y)}}{G(y)}.  \label{51}
\end{equation}%
Based on the definition of surface gravity, the Hawking temperature can be
written as%
\begin{equation}
T_{h}=\frac{l^{2}}{4\pi R}\frac{1+\nu }{\sqrt{\lambda \nu }}\sqrt{\frac{%
1-\lambda }{1+\lambda }}.  \label{52}
\end{equation}%
The energy flux and charge flux from the 5D neutral black ring are%
\begin{equation}
\dot{E}=\frac{\pi T_{h}^{2}}{12}+\frac{l^{2}}{4\pi }(\frac{F(y_{h})}{CR{%
(1+y_{h})}})^{2},  \label{53}
\end{equation}%
\begin{equation}
\dot{I_{l}}=-\frac{l^{2}A_{t}(y_{h})}{2\pi }=\frac{l^{2}}{2\pi }\frac{%
F(y_{h})}{CR{(1+y_{h})}}.  \label{54}
\end{equation}%
We find the energy flux and charge flux equal to Hawking radiation flux and
gauge flux respectively, which means that thermal radiation and particles
from black ring also can be transported by a 1D quantum channel. The thermal
radiation and particles can be transported from black ring mainly stems from
that it also owns an event horizon, which can be treated as the thermal
source.

\section{Discussion and conclusion}

Flows of energy and charge from the higher dimensional Reissner-Nordstr$%
\ddot{o}$m-de Sitter black hole, Kerr-Newman black hole, and 5D black ring
have been studied via the Landauer transport model. Though these black holes
have different topological geometry, but all of them own horizons, which can
emit Hawking radiation with Hawking temperature, hence the horizons and its
environments can be connected by 1D quantum channel as two thermal
reservoirs and the Landauer transport model is valid. We find the energy
current of fermions and bosons transported by a 1D quantum system can be
divided into two independent components: the flux of charge, which is only
related to the particles entirely, and the flux of thermal radiation, which
are determined by the temperature of thermal reservoir irrespective to the
number of particles. The reason for particle statistics independence of
Landauer transport model maybe stems from that the mutual cancellation of
the group velocity and density of states enter the current formulae \cite%
{8,10,10-1,10-2,10-3}.

It should be stressed that in this paper, the Landauer transport model
provides not the method to calculate Hawking radiation but the probable way
how Hawking radiation runs, namely by a 1D quantum tunnel.

\begin{acknowledgments}
This research is supported by the National Natural Science Foundation of
China (Grant Nos. 10773002, 10875012). It is also supported by the
Scientific Research Foundation of Beijing Normal University under Grant No.
105116.
\end{acknowledgments}


\begin{thebibliography}{99}
\bibitem{1} S. W. Hawking, Nature 248 (1974) 30.

\bibitem{2} J. B. Hartle, S. W. Hawking, Phys. Rev. D 13 (1976) 2188.

\bibitem{2-1} G. Gibbons, S. W. Hawking, Phys. Rev. D 15 (1977) 2752.

\bibitem{3} S. M. Christensen, S. A. Fulling, Phys. Rev. D 15 (1977) 2088.

\bibitem{4} S. P. Robinson, F. Wilczek, Phys. Rev. Lett. 95 (2005) 011303.

\bibitem{4-1} R. Banerjee, S. Kulkarni, Phys. Rev. D 77 (2008) 024018.

\bibitem{4-2} S. Iso, T. Morita, H. Umetsu, Phys. Rev. D 76 (2007) 064015.

\bibitem{4-3} Z. Xu, B. Chen, Phys. Rev. D 75 (2007) 024041.

\bibitem{4-4} M. R. Setare, Euro. Phys. J. C 49 (2006) 865.

\bibitem{4-5} S. Iso, T. Morita, H. Umetsu, J. High. Energy Phys. 04 (2007)
068.

\bibitem{4-6} Q. Q. Jiang, S. Q. Wu, X. Cai, Phys. Lett. B 651 (2007) 58.

\bibitem{4-7} K. Xiao, W. B. Liu, H. B. Zhang, Phys. Lett. B 647 (2007) 482.

\bibitem{4-8} Q. Q. Jiang, S. Q. Wu, X. Cai, Phys. Rev. D 75 (2007) 064029.

\bibitem{4-9} S. Q. Wu, J. J. Peng, Class. Quant. Grav. 24 (2007) 5123.
hep-th/0706.0983.

\bibitem{5} G. W. Gibbons, M. J. Perry, Phys. Rev. Lett. 36 (1976) 985.

\bibitem{5-1} G. W. Gibbons, M. J. Perry, Proc. Roy. Soc. Lond. A 358 (1978)
467.

\bibitem{6} M. K. Parikh, F. Wilczek, Phys. Rev. Lett. 85 (2000) 5042.

\bibitem{6-1} J. Zhang, Z. Zhao, Nucl. Phys. B 725 (2005) 173.

\bibitem{6-2} J. Zhang, Z. Zhao, JHEP 0510 (2005) 055.

\bibitem{6-3} W. B. Liu, Phys. Lett. B 634 (2006) 541.

\bibitem{6-4} S. Z. Yang, Chin. Phys. Lett. 22 (2005) 2492.

\bibitem{6-5} E. C. Vagenas, Phys. Lett. B 503 (2001) 399.

\bibitem{6-6} Q. Q. Jiang, S. Q. Wu, X. Cai, Phys. Rev. D 73 (2006) 064003.

\bibitem{6-7} S. Q. Wu, Q. Q. Jiang, JHEP 0603 (2006) 079.

\bibitem{6-8} M. K. Parikh, Int. J. Mod. Phys. D 13 (2004) 2355.

\bibitem{6-9} S. W. Zhou, W. B. Liu, Mod. Phys. Lett. A 24 (2009) 2099.

\bibitem{6-10} X. X. Zeng, Mod. Phys. Lett. A 24 (2009) 625.

\bibitem{7} J. M. Maldacena, Adv. Theor. Math. Phys. 2 (1998) 231.

\bibitem{7-1} L. Susskind, J. Math. Phys. 36 (1995) 6377.

\bibitem{8} P. D. Nation, M. P. Blencowe, F. Nori, Landauer transport model
for Hawking radiation from a black hole, arXiv: gr-qc/1009.3974 (2010).

\bibitem{9} Y. Imry, R. Landauer, Rev. Mod. Phys. 71 (1999) 306.

\bibitem{9-1} Y. Imry, Introduction to Mesoscopic Physics, 2nd ed. (Oxford,
2008).

\bibitem{10} U. Sivan, Y. Imry, Phys. Rev. B 33 (1986) 551.

\bibitem{10-1} L. G. C. Rego, G. Kirczenow, Phys. Rev. Lett. 81 (1998) 232.

\bibitem{10-2} M. P. Blencowe, Phys. Rev. B 59 (1999) 4992.

\bibitem{10-3} K. Schwab, Nature 404 (2000) 974.

\bibitem{11} J. B. Pendry, J. Phys. A 16 (1983) 2161.

\bibitem{12} W. G. Unruh, Phys. Rev. D 14 (1976) 870.

\bibitem{13} S. Iso, H. Umetsu, F. Wilczek, Phys. Rev. D 74 (2006) 044017.

\bibitem{14} R. Banerjee, S. Kulkarni, Phys. Rev. D 79 (2009) 084035.

\bibitem{14-1} L. Alvarez-Gaume, E. Witten, Nucl. Phys. B 234 (1984) 269.

\bibitem{14-2} W. A. Bardeen, B. Zumino, Nucl. Phys. B 244 (1984) 421.

\bibitem{14-3} H. Banerjee, R. Banerjee, Phys. Lett. B 174 (1986) 313.

\bibitem{14-4} R. Bertlmann, Anomalies in quantum field theory. (Oxford:
Oxford Science, 2000).

\bibitem{15} R. Landauer, IBM J. Res. Dev. 1 (1957) 223.

\bibitem{15-1} R. Landauer, Philos. Mag. 21 (1970) 863.

\bibitem{16} Y. S. Wu, Phys. Rev. Lett. 73 (1994) 922.

\bibitem{17} L. G. C. Rego, G. Kirczenow, Phys. Rev. B 59 (1999) 13080.

\bibitem{18} L. D. Landau, E. M. Lifshitz, Statistical physics, 3rd ed.
(Oxford: Pergamon Press, 1980).

\bibitem{19} E. Witten, Adv. Theor. Math. Phys. 2 (1998) 253.

\bibitem{20} R. P. Kerr, Phys. Rev. Lett. 11 (1963) 237.

\bibitem{20-1} E. T. Newman, A. I. Janis, J. Math. Phys. 6 (1965) 915.

\bibitem{21} R. Emparan, JHEP 03 (2004) 064.
\end{thebibliography}
\end{document}